# Is modern science evolving in the wrong direction?

Klaus Jaffe
American Center for Strategic Studies
CEE-IAEAL, Universidad Simón Bolívar
Caracas, Venezuela
kjaffe@usb.ve

**Abstract:** The present politically correct consensus is that increased international exchange of scientific insight, knowledge, practitioners and skills at the global level brings significant benefits to all. The quantifiable scientometric changes during the last decade, however, suggest that many areas of knowledge are evolving in the opposite direction. Despite an increase during the last decade of the numbers of journals and academic articles published, increases in the number of citations the published articles receive, and increases in the number of countries participating; important parts of the academic activity are becoming more nationalistic. In addition, international collaboration is decreasing in several subject areas, and in several geographic regions. For example, countries in Asia are becoming scientifically more isolated; and academics working in the humanities in all the regions of the world are very nationalistic and are becoming more so. The precise consequences of this dynamics are difficult to predict, but it certainly will have reverberations beyond academia. The tendency of the humanities to become more provincial will certainly not help in reducing international conflicts arising from poor understanding of cultural differences and of diverging sociopolitical world views. More and better data on these trends should give us a better understanding for eventually improving academic policies worldwide.

**Keywords:** International Collaboration, National, Science, Humanities, Policy

# ¿Está evolucionando la ciencia moderna en la dirección equivocada ?

Klaus Jaffe
Centro de Estudios Estratégicos
CEE- IAEAL , Universidad Simón Bolívar
Caracas, Venezuela
kjaffe@usb.ve

**Resumen**: El consenso actual considerado políticamente optimo es que el aumento del intercambio científico internacional incluyendo conocimiento, información , profesionales y habilidades técnicas trae beneficios significativos para todos. Los cambios bibliométricos cuantificables durante la última década, sin embargo , sugieren que muchas áreas del conocimiento están evolucionando en la dirección opuesta . A pesar del aumento en esa década del número de revistas y artículos académicos publicados , el aumento en el número de citas que reciben los artículos publicados, y del aumento en el número de países participantes, partes importantes de la actividad académica son cada vez más nacionalistas. Además, la colaboración internacional está disminuyendo en varias áreas temáticas, y en varias regiones geográficas. Por ejemplo , los países de Asia están cada vez más aislados científicamente, y académicos que trabajan en las humanidades en todas las regiones del mundo son muy nacionalistas y se están convirtiendo aun más centrados en su pais. Las consecuencias precisas de esta dinámica son difíciles de predecir , pero sin duda tendrá repercusiones más allá de los círculos académicos . La tendencia de las humanidades en ser una actividad cada vez mas local, sin duda no ayuda en la reducción de los conflictos internacionales derivados de la mala comprensión de las diferencias culturales y de la divergencia de puntos de vista socio-políticos. Más y mejores datos sobre estas tendencias deberían darnos una mejor comprensión de este fenómeno, que nos permitan diseñar políticas académicas que beneficies a largo plazo a todos los países del mundo .

**Palabras clave** : Colaboración Internacional, Nacionalismo, Ciencias , Humanidades, Política

## Introduction

The emergence of empirical science was the foundation for the revolution in technological expertise that triggered the industrial revolution, which marked the world economy during the last few centuries (Jaffe, 2009). From its beginnings, science was based on international collaboration. Yet science has changed since the days of Galileo, Newton and the foundation of the Academia dei Lincey in 1603 and the Royal Society in 1660. The way we value and promote the different sciences affects our economies (Jaffe, 2005). Knowing how modern sciences are changing and how they will look in future is essential if we want to understand and manage future economic developments.

Different sciences and scientific disciplines cultivate different values and attitudes and show differences in quantifiable characteristics (Fanelli, 2010; Jaffe et al., 2010; Filipi et al. 2012; Fanelli & Glänzel 20l3). We also know that the development of different scientific disciplines has different effects on economic growth. For example, the subject areas with the largest relative number of publication in wealthy countries today are neuroscience and psychology; investment in these areas however does not produce economic growth in less developed countries. In contrast, middle income countries that give more value to basic natural science in a given time period show faster economic growth in the following years (Jaffe et al. 2013b), showing that the structure of the national scientific ecosystem affects society. Additionally, countries whose researchers are less provincial and cite more works from countries different to theirs (have fewer country self-citations) are also those whose scientists produce relatively lower numbers of author self-citations. These countries are the ones producing scientific papers with higher overall citation impact (Jaffe 2011).

A recent report by the Royal Society of London (Royal Society 2011), stresses that international collaboration improves the quality of the scientific papers produced, that Science is increasingly global and multipolar; that the scientific world is becoming increasingly interconnected, with international collaboration on the rise forming networks that span the globe. The report emphasizes a future for exchange of scientific insight, knowledge and skills, with a change of focus of science from the national to the global level that will bring significant benefits to all.

The question posed in the present paper is if modern academia is actually evolving in this direction. Improving our understanding of the changing patterns of science, scientific institutions and academic collaboration, is essential to identify the opportunities and benefits of international collaboration, to consider how they can best be realized, and how they can be harnessed to tackle global problems more effectively.

## Methods

In order to answer these questions, changes in available quantitative scientometric variables were computed from 21135 journals, for 20 different subject areas, grouped by Scopus, and reported by SCImago (2007). The statistical methods used (non-parametric correlations using Statistica) were shown to be robust and their limitations were analyzed in detail before (Jaffe et al., 2013a). The time period chosen to sample the data, from 1999 to 2011, guarantees a more or less uniform bibliometric methodology and enough time for the most recent data pools to have retrieved most of the corresponding data (some journal issues appear years after their listed publication year). Without exceptions, the data used here for 1999 and 2011 were always at the extremes of the range studies, with values for the intermediate years falling between these two extremes. The variables analyzed are summarized in Table 1

**Table 1**: Quantitative variables used

| | |
|---|---|
| **IC** | International Collaboration: Proportion of document with affiliations from more than one country |
| **Journals** | Number of Journals tracked by Scopus in a given subject category |
| **Doc/Jour** | Number of citable documents per journal in a given subject category |
| **Countries** | Number of countries reported in the addresses of the authors of the papers in that subject category |
| **Ref/Doc** | Number of references in the papers published in that subject category |
| **Cit/Doc** | Number of citations received during the following 3 years after publication by papers in that subject category |
| **CSC** | Level of provinciality or degree of country-self citation measured as the proportion of citations from the same country as the source paper. Country self-citations include author-self citations. |

## Results

The results from this analysis revealed various features that differ between subject areas. For example, different subject areas vary in the number of papers their journal publishes. During the 12 year period studied, all subject areas increased the number of their journals, the average number of papers published in these journals, and the number of citation per article published. The relative difference between subject areas in the number of journals and average number of papers per journal changed little during this period (Figure 1).

**Figure 1**: Average number of papers per journal (Doc/Jour) plotted against the total number of journals registered by Scopus (Journals) for each of the subject area for two different years. The size of bubbles is proportional to the average total number of citations for papers published 3 years earlier (Cit/Doc) as computed by SCImago (2009).

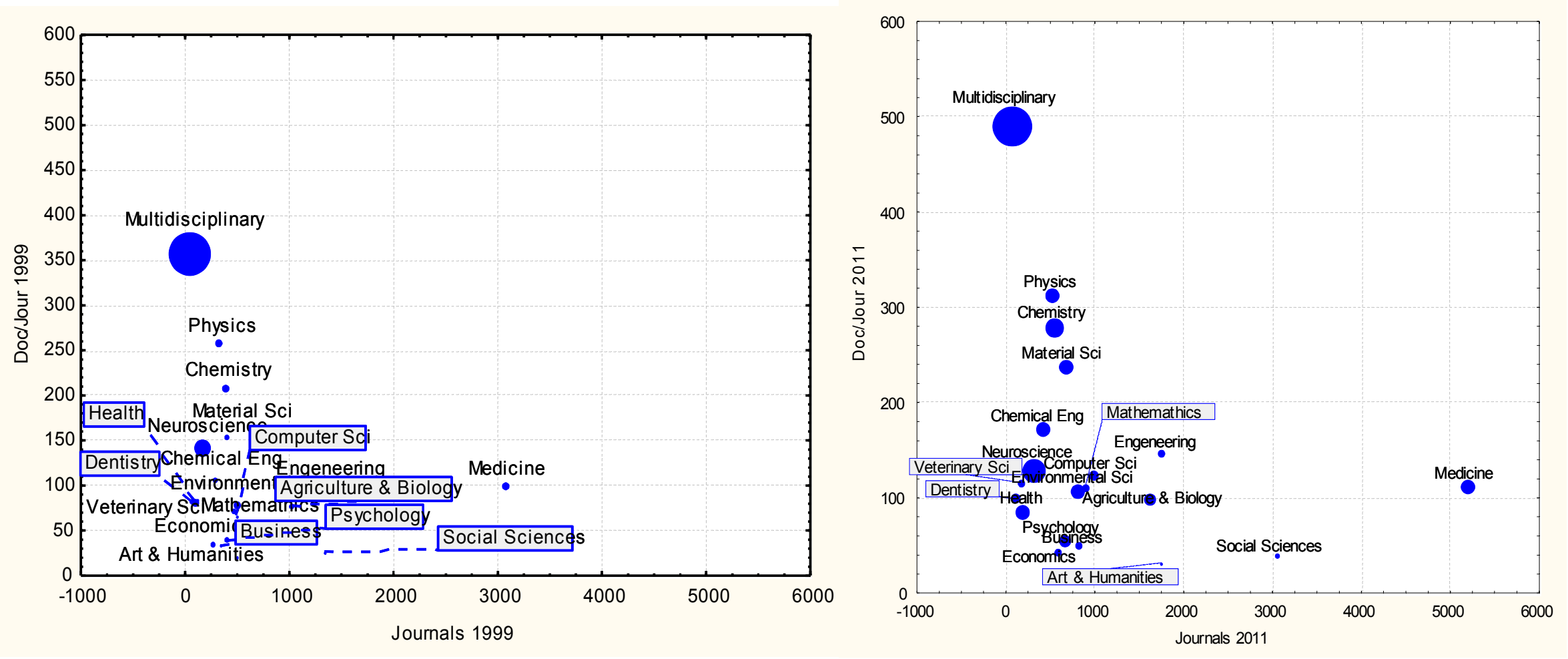

**Table 2**: Gamma correlations of the relationship between different bibliometric measures taken in the same year for the different subject areas. (For example, the correlation between the number of countries reported in the publications and the number of journals in that subject area in 1999 is 0.68)

| | Journals | p | Doc/Jour | p |
|---|---|---|---|---|
| **Countries 1999** | **0.65** | **0.002** | 0.23 | 0.33 |
| **Countries 2011** | **0.62** | **0.003** | 0.19 | 0.42 |
| **Ref/Doc 1999** | 0.19 | 0.42 | **-0.45** | **0.0.5** |
| **Ref/Doc 2011** | -0.09 | 0.69 | -0.13 | 0.6 |
| **Cit/Doc 1999** | -0.40 | 0.08 | **0.66** | **0.001** |
| **Cit/Doc 2011** | -0.40 | 0.08 | **0.64** | **0.003** |
| **CSC 1999** | 0.09 | 0.69 | **-0.53** | **0.01** |
| **CSC 2011** | **0.46** | **0.04** | -0.25 | 0.28 |

A finer quantitative statistical analysis of the scientometric differences between the 20 subject areas studied is presented in Table 2. This analysis shows that in the two time periods studied, the number of journals in each subject area correlated with the number of countries in which the scientist publishing the papers came from. That is, the more journals the subject areas possessed, the more diverse the countries that had active scientists in the subject category. Subject areas with journals with high number of publications (Pub/Jour)), published papers with high citation rates and relatively lower country-self-citation rates (CSC). That is, subject areas with high average citation rates published more papers per journal, and those papers had relatively lower country-self-citations.

**Figure 2**: Changes in country self-citation rates (CSC) between 1999 and the year 2011 plotted against changes in citation impact (Cit/Doc) during the same time period. The size of bubbles is proportional to Cit/Doc in 1999. The line shows the linear regression.

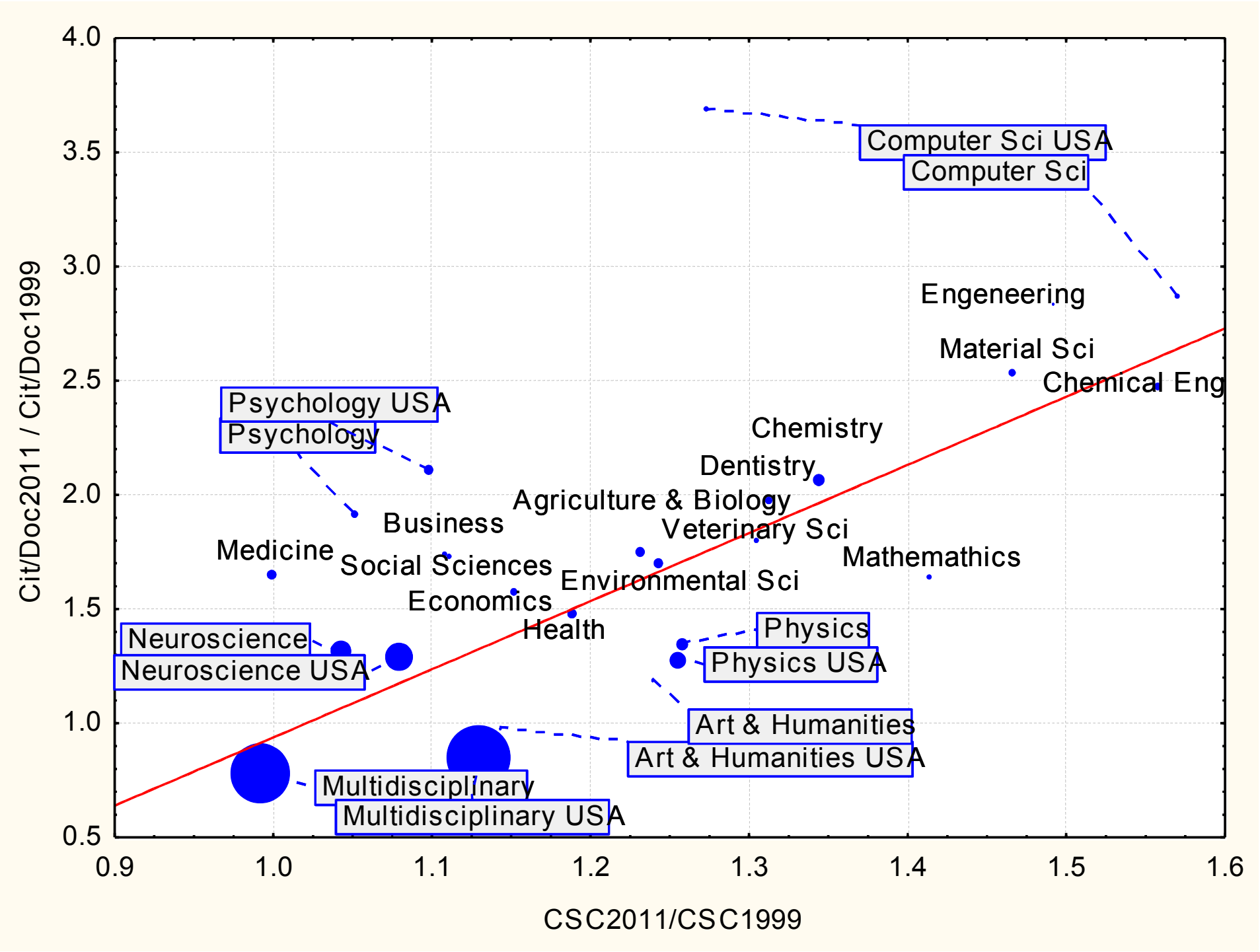

**Table 3**: Gamma correlations comparing the changes in bibliometric variables of the subject areas from 1999 to 2011, expressed as the ratio of the variable for 2011/1999.

*** indicate correlations with p<0.05 and ** p<0.01**

| | Ratio Country | Ratio Cit/Doc | Ratio Ref/Doc | Ratio CSC | Ratio Doc/Jour |
|---|---|---|---|---|---|
| **Ratio Journals** | **0.50**** | -0.30 | -0.25 | -0.21 | 0.05 |
| **Ratio Country** | └ | -0.30 | -0.17 | -0.17 | -0.03 |
| **Ratio Cit/Doc** | | └ | 0.18 | **0.52**** | **0.38**** |
| **Ratio Ref/Doc** | | | └ | 0.18 | -0.04 |
| **Ratio CSC** | | | | └ | **0.42**** |

If we focus on the change during the 12 year period, we detect a pattern of statistically significant correlations (Table 3) that shows that increases in journal numbers (Journals) correlated with increases in the number of countries participating (Countries). It also showed that the change in citation impact (Cit/Doc) correlated positively with the change in the degree of country-self-citation (CSC) and with number of documents published per journals (Doc/Jour). Increases in country-self-citations (CSC) correlated with increases in citation impact (Cit/Doc) and with increases in number of documents per journal (Doc/Jour). That is, subject areas with many new journals, publishing articles with fewer references per paper, and had higher county-self-citations.

These changes in time are presented graphically in Figure 2 which shows that, except for multidisciplinary sciences and art and humanities, all subject areas increased their citation impact (Cit/Doc 2011 / Cit/Doc1999), and to a lower degree, also their country-self-citations (CSC2011/CSC1999). The subject areas with the largest increase in citations and the ones with the largest increase in country-self-citations were the ones which had the lowest citation impact in 1999.

These trends were similar for world wide data and data for the USA only, the country with the largest scientific activity in the world (19.5% of the total in 2011). The USA however showed some remarkable differences in some subject areas. For example in Computer Science, the increase in country-self-citations was much larger in the rest of the world compared to the USA, but that was due to the fact that the USA had the highest country-self-citations, together with Iran, in 1999.

The data then shows that although all subject areas increased their scientific activity (Fig 1), they also increased their CSC (Fig 2), except in multidisciplinary sciences (with a low CSC). The arts and humanities with one of the highest CSC, slightly increased CSC further although. That is, all academic subject areas, except multidisciplinary science, are becoming more nationalistic or provincial. Multidisciplinary science seems to be particular in that it has few journals, maintains a high impact factor and was the only subject area that decreased its CSC.

A separate analysis showed that the trend in International Collaboration (IC) was very heterogeneous between the geographical regions studied (Figure 3). The Pacific region had the third highest IC in 1999 and increased to first place in 2011. Western Europe and North America also increased their IC reaching second and fifth place respectively in 2011. But Latin America Eastern Europe and the Asian Region reduced their IC between 1999 and 2011. The Asian Region had the worst IC in both years.

**Figure 3:** Percentage of papers with coauthors from different counties (IC) for seven different geographical regions in 1999 and 2011

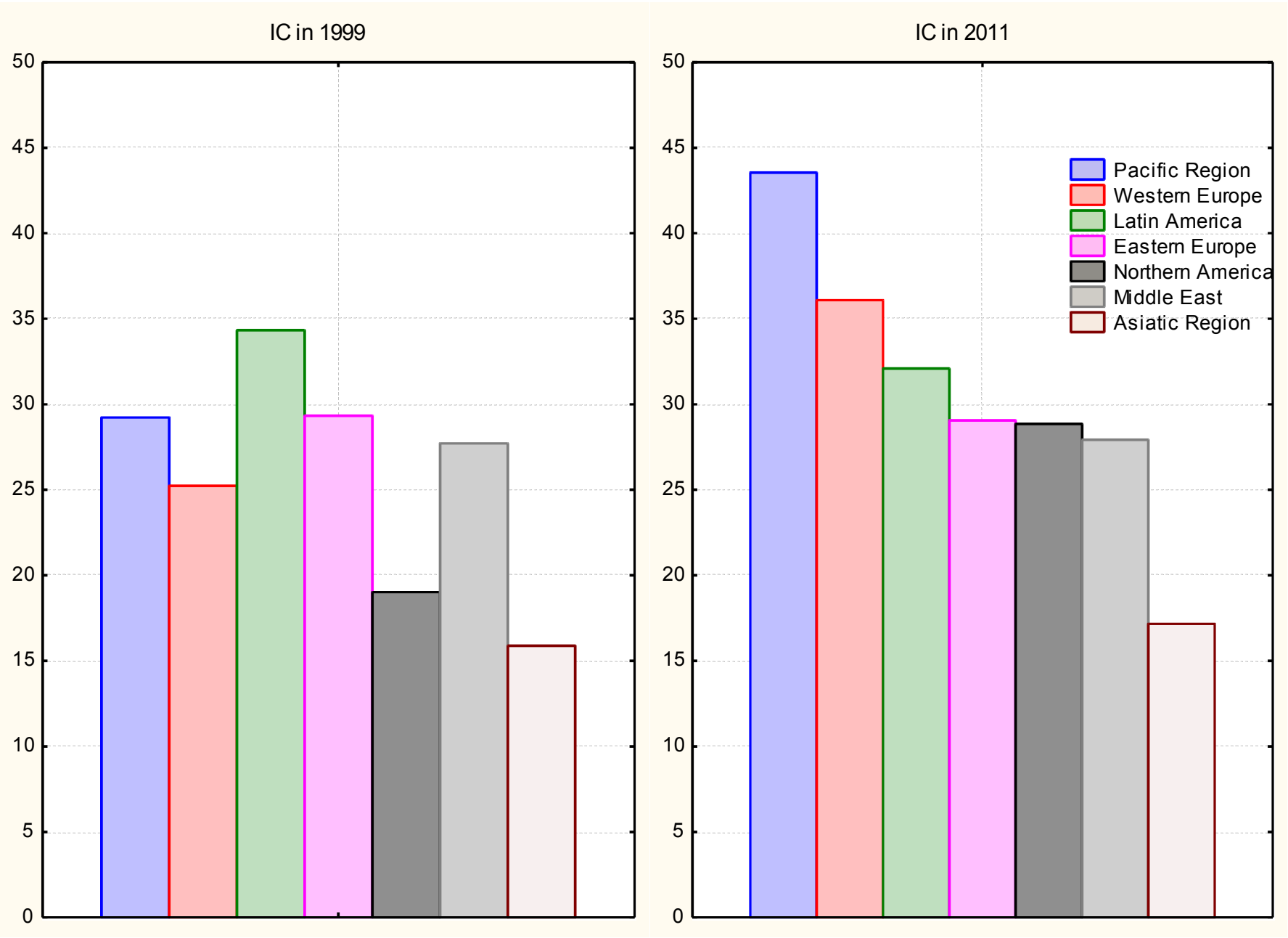

In Figure 4 we compare the region with the second highest rate of IC, Western Europe, with Asia showing the lowest IC. In both regions, Humanities had the lowest IC whereas Multidisciplinary Sciences and Physics in Western Europe, and Economics and Psychology in Asia, were the subject areas with the highest IC.

The different subject areas showed large variations in their rate of change in IC during the 12 years studied (Figure 5). IC in Arts & Humanities decreased in 2011 compared to 1999 in all four geographical regions, even considering that IC in this subject area was the lowest in 1999. IC among engineers and computer scientists increased in all four regions in that period. In Asia, IC in Business, Medicine and Chemistry increased the most, whereas in Western Europe it was IC in Engineering, Medicine and Psychology. In general, the pattern for North America was similar to that of Western Europe (see SM3 supplementary information). Subject areas that showed large IC in North America and Western Europe in 1999, where the ones with the smallest further expansion in IC in Western Europe as assessed in 2011, except Arts & Humanities (Fig 5).

**Figure 4:** Percentage of International Collaboration (IC) in different subjected areas in Asia plotted against IC in Western Europe. The size of the bubbles is proportional to the citation impact (Cit/Doc) in of the subject area in 2011.

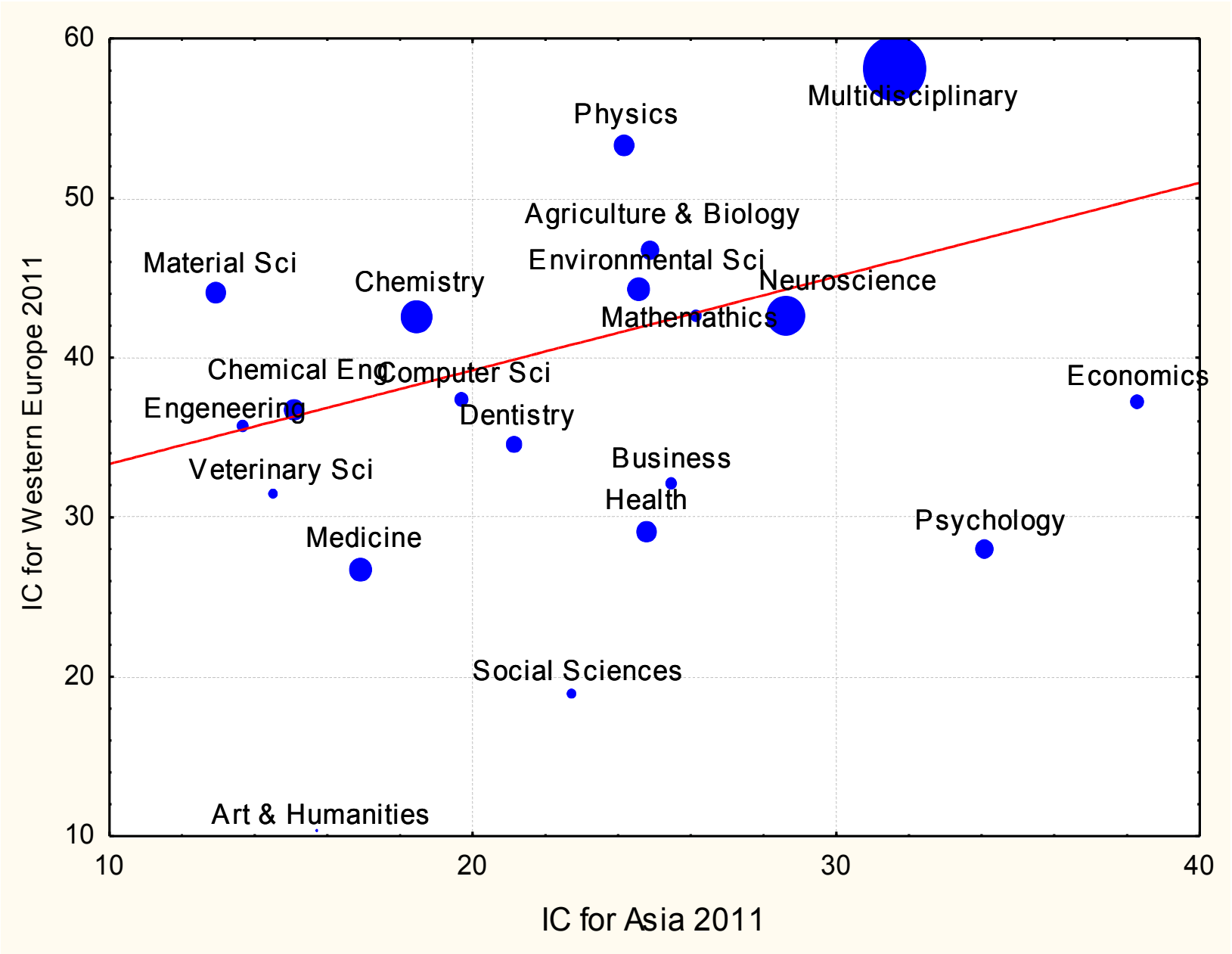

**Figure 5:** Rate of change in International Collaboration (IC) expressed as IC in 2011 / IC in 1999. This rate for different subjected areas in Asia is plotted against that in the same subject areas Western Europe. The size of the bubbles is proportional to IC of the subject area in the North America in 1999.

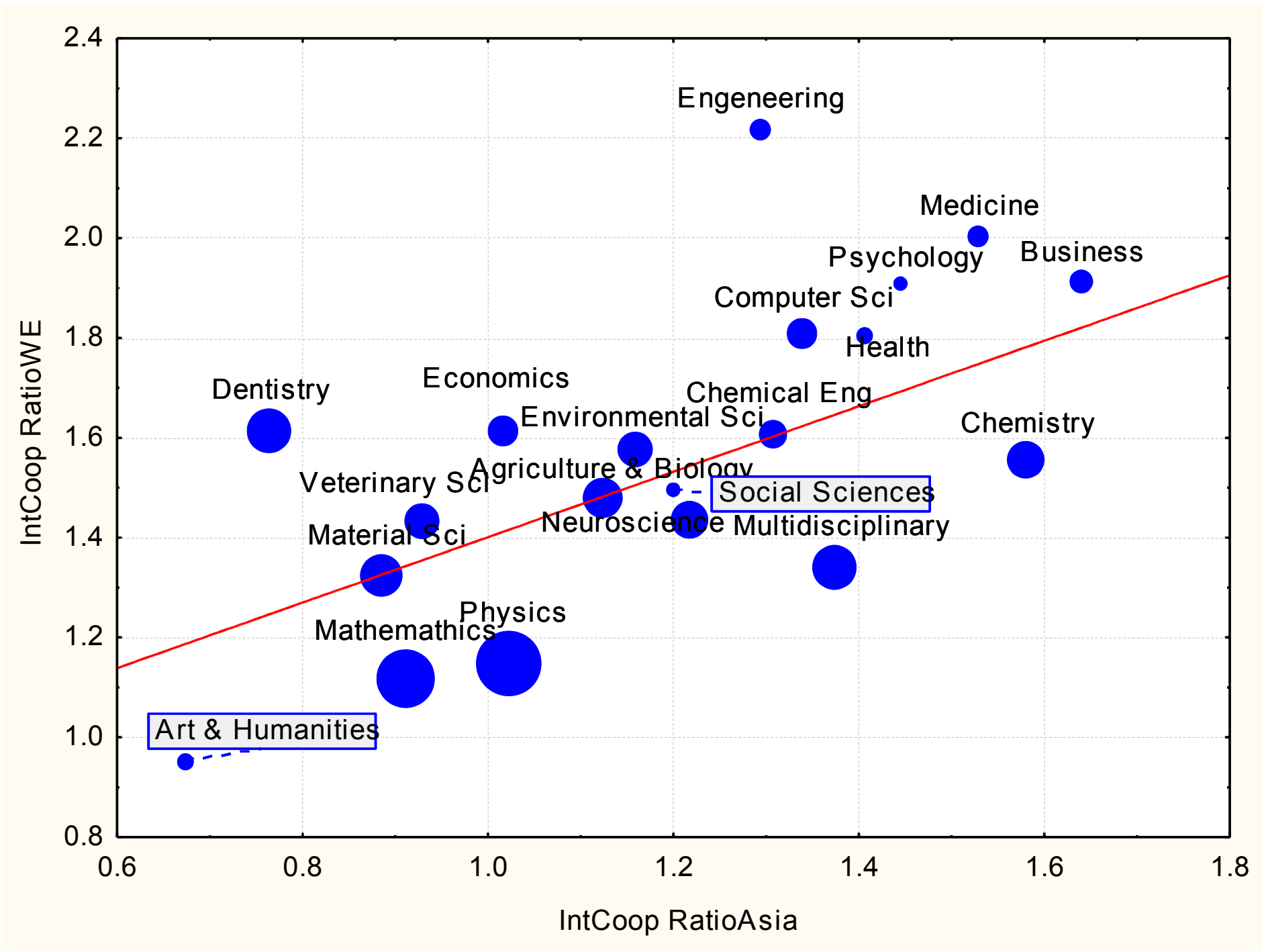

## Discussion

The results show that bibliometric trends, reported some decades ago (Frame & Carpenter 1979; Luukkonen 1992; Persson et al., 2004), continue to be valid. That is, the more basic the field of knowledge, the greater the proportion of international collaboration; and the larger the national scientific enterprise, the smaller the proportion of international collaboration.

The data reflect the many new online journals started by scientists in an increasing number of countries during the last decade, increasing the overall number of journals and the number of countries participating in the scientific activity of the different subject areas. Some subject areas, such as Multidisciplinary Science, have fewer journals publishing more articles, whereas others, such as Medicine, have many journals, each one publishing fewer articles. This difference was maintained during the time period studied. Thus, some characteristics of subject areas seem to be resilient to change despite the large increment in journals.

The surprising result of this study is the tendency in the last decade for several areas to increase their country self-citations and to decrease their International Collaborations. That is, several subject areas and several geographic areas are become more provincial, where some disciplines and countries become more isolated respect to scientific activity, a trend completely opposite to that recommended by the Royal Society (2011). This policy document concluded that international "collaboration brings significant benefits, both measurable (such as increased citation impact and access to new markets), and less easily quantifiable outputs, such as broadening research horizons. The facilitation of collaboration, therefore, has a positive impact not only on the science conducted, but on the broader objectives for any science system (be that enhancing domestic prosperity or addressing specific challenges)".

International Collaboration was highest and continued to expand faster in subject areas related to basic natural science, and was lower and increased less in areas of applied sciences and the humanities. This trend is consistent with the advice of some politically influential economists (Sachs, 2005; Hausmann et al 2011 for example) who recommend nations to focus on applied research relevant to their specific national problems. They advise countries to plan their scientific activity to achieve practical. It is thus very interesting to observe the economic success of countries not following this recommendation, but that invest relatively more in basic natural science (Jaffe et al., 2013b). The present analysis favors the view that focusing on fomenting basic research might foment more international collaborations, because we know that applied research foments less international collaboration than basic research in natural sciences (Frame & Carpenter 1979). International Collaboration in basic research spills over other areas, strengthening competitive advantages developed through international competition, unleashing synergies that stimulate economic growth (Stutz & Barney, 2007). Politicians and humanist in general believe in the unbound capabilities of our mind to plan our future. Basic science though acknowledges that the future is often unpredictable, and works fomenting synergies and favoring creative serendipity.

Clearly, the health of all parts of the present international academic system is not robust. Several subject areas are becoming less international, and several countries are becoming academically more isolated. The consequence of this dynamics is difficult to predict, but it will have reverberations beyond science, and thus should be studied more carefully. The tendency of the humanities to become more isolated will certainly not help in reducing international conflicts arising from poor understanding of cultural differences and of diverging sociopolitical world views. More and better data on these trends should give us a better understanding for eventually improving policies fomenting knowledge in general and science and humanities in particular worldwide.

**References**

Fanelli D, Glänzel W (2013) Bibliometric Evidence for a Hierarchy of the Sciences. PLoS ONE 8(6): e66938. doi:10.1371/journal.pone.0066938

Fanelli D. (2010). ‘‘Positive’’ Results Increase Down the Hierarchy of the Sciences PLoS ONE 5(4): e10068. doi:10.1371/journal.pone.0010068

Filipi N. Silva, Francisco A. Rodrigues, Osvaldo N. Oliveira Jr, Luciano da F. Costa (2012). Quantifying the interdisciplinarity of scientific journals and fields. arXiv:1203.4807v1

Frame J.D and Carpenter M P (1979) International Research Collaboration. Social Studies of Science 9: 481-497.

Hausmann R, Hidalgo CA, Bustos S, Coscia M, Chung S, Jimenez J, Simoes A, Yıldırım MA (2011) The Atlas of Economic Complexity: mapping paths to prosperity.

Jaffe K (2005) Science, religion and economic development. Interciencia 30: 370–373.

Jaffe K (2009) What is Science? An interdisciplinary perspective. University Press of America, USA.

Jaffe K, Caicedo M, Manzanares M, Gil M, Rios A, et al. (2013) Productivity in Physical and Chemical Science Predicts the Future Economic Growth of Developing Countries Better than Other Popular Indices. PLoS ONE 8(6): e66239. doi:10.1371/journal.pone.0066239

Jaffe K, Rios A, Florez A (2013) Statistics shows that economic prosperity needs both high scientific productivity and complex technological knowledge, but in different ways. Interciencia 38: 150–156.

Jaffe K. (2011) Do countries with lower self-citation rates produce higher impact papers? Or, does humility pay. Interciencia 36: 696-698.

Jaffe, K., Florez, A., Grigorieva, V. Masciti, M., Castro, I. (2010) Comparing skills and attitudes of scientists, musicians, politicians and students, Jaffe, K., Florez, A., Grigorieva, V. Masciti, M., Castro, I. Interciencia 35, 545-552.

Luukkonen T (1992) Understanding Patterns of International Scientific Collaboration. Science Technology Human Values. 17: 101-126.

Persson O, Flanzel W. Danell R. (2004) Inflationary bibliometric values: The role of scientific collaboration and the need for relative indicators in evaluative studies. Scientometrics 60: 421-432.

Royal Society (2011) Knowledge, Networks and Nations: Global Scientific Collaboration in the 21st Century. RS Policy document 03/11. London, UK: http://royalsociety.org/uploadedFiles/Royal_Society_Content/Influencing_Policy/Reports/2011-03-28-Knowledge-networks-nations.pdf

Sachs J (2005) The end of Poverty. Penguin Press 416 pp.

SCImago (2007) SJR — SCImago Journal & Country Rank. Retrieved from 14-8 to 12-9, 2013 from http:www.scimagojr.com

Stutz, F.P. Barney, W (2007). *The world economy: resources, location, trade and development* (5th ed. ed.). Upper Saddle River: Pearson.

## Figure 1:

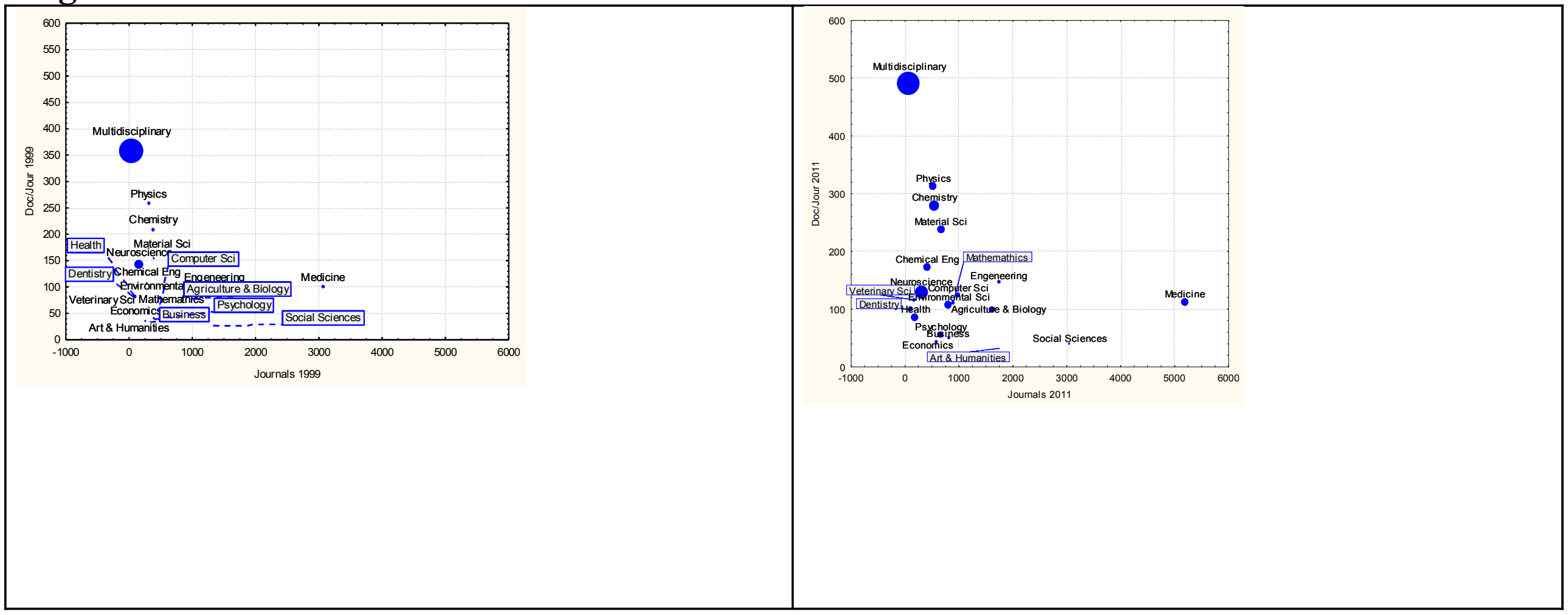

## Figure 2:

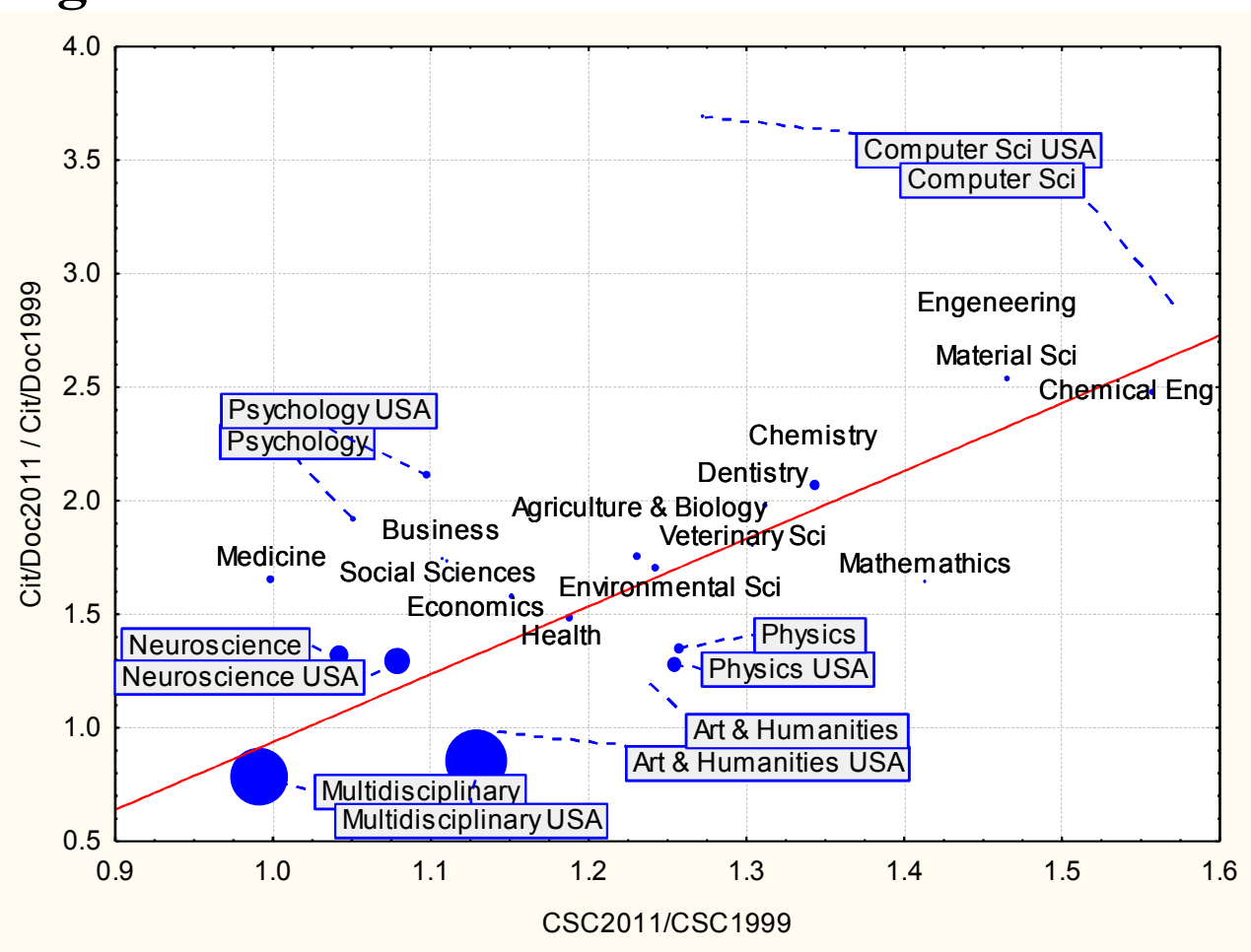

**Figure 3:**

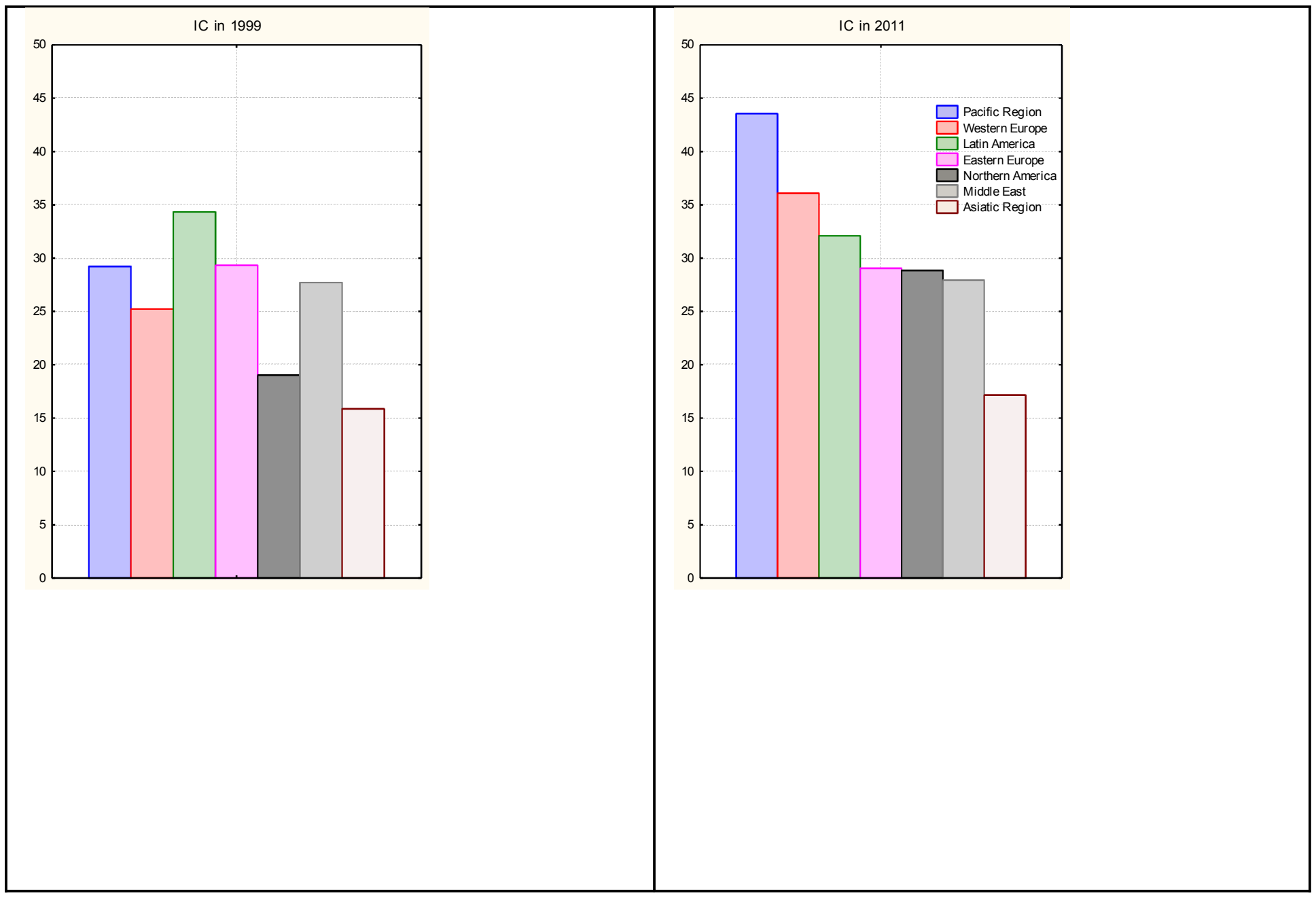

**Figure 4:** .

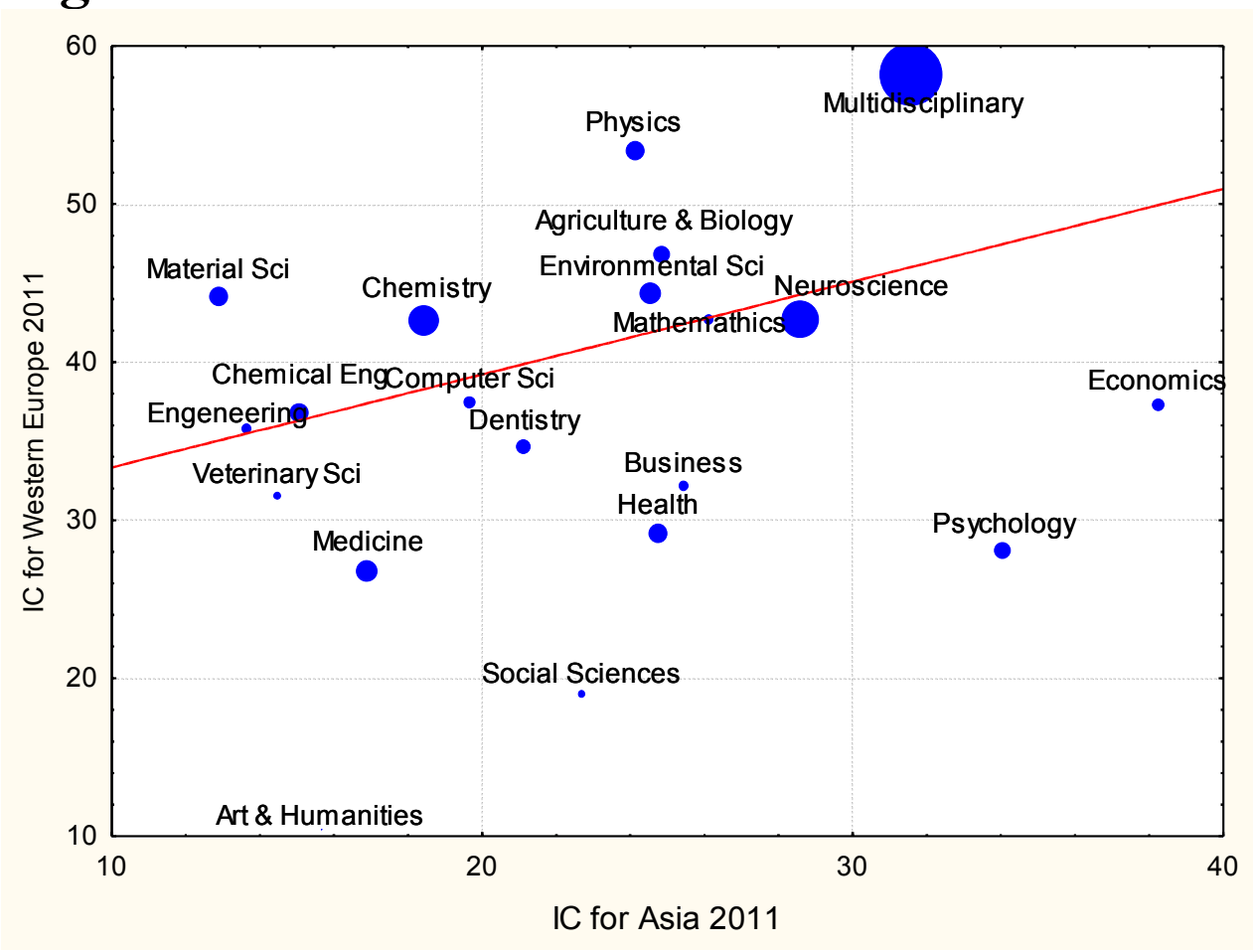

# Figure 5:

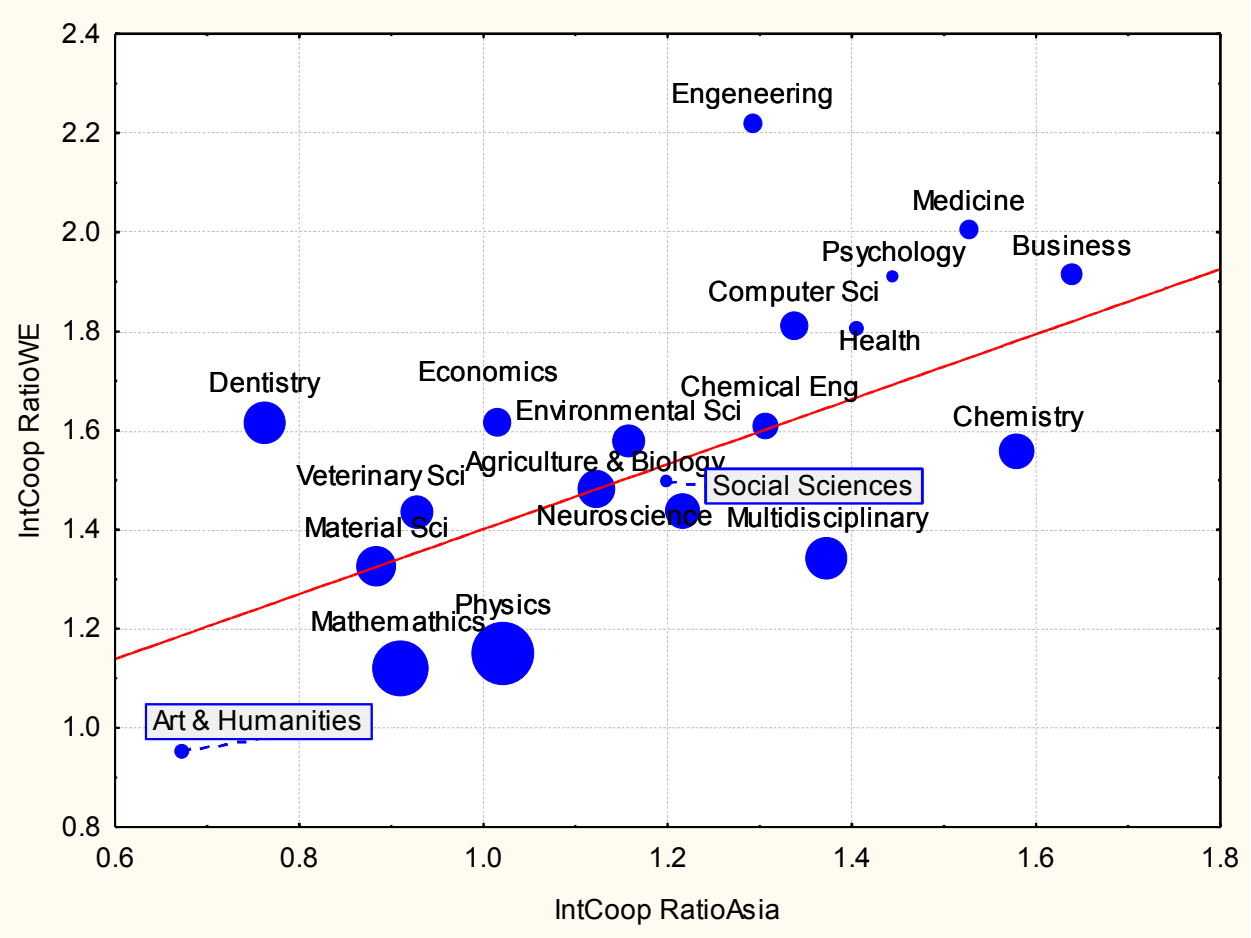

**Significance:** This bibliometic study shows that most areas of science are becoming less international regarding their use of references and that the humanities and other academic areas have little international collaborations and further reduced them in the last decade. These trends are alarming in that they oppose recommended best practice in science (by The Royal Society, American Academies of Science, AAAS and others) with potential unfortunate consequences affecting society beyond academia. Publication of hard data unveiling these dangerous trends is essential if we want to design science policies to correct them.